\documentclass{icrc2009}

\usepackage{graphicx}   % for including figures
\usepackage{caption}    % for captions
\usepackage[font=footnotesize]{subfig} % subfig.sty for a double column floating figure using two subfigures
\usepackage{fixltx2e}
\usepackage{url}

\newcommand{\shorttitle}[1]%
{\markboth{Proceedings of the 31\MakeLowercase{$^{st}$} ICRC, {\L}\'{o}d\'{z} 2009}{#1} }
\newcommand{\etal}{\MakeLowercase{\textit{et al. }}} % "et al."

\begin{document}
%
% ICRC 2009 Lodz proceedings
%
\title{Technical Performance of the MAGIC Telescopes}

\author{\IEEEauthorblockN{Juan Cortina\IEEEauthorrefmark{1},
                          Florian Goebel\IEEEauthorrefmark{2},
			  Thomas Schweizer\IEEEauthorrefmark{2},
                          for the MAGIC Collaboration}
                            \\
\IEEEauthorblockA{\IEEEauthorrefmark{1}Institut de Fisica d'Altes Energies, Cerdanyola del Valles, E-08193 Spain}
\IEEEauthorblockA{\IEEEauthorrefmark{2}Max-Planck-Institut f\"ur Physik, D-80805 M\"unchen, Germany}
}

% please write the presenter's name and short title (3-4 words maximum)
%    which will appear at the header of the even pages.
\shorttitle{Juan Cortina \etal Technical MAGIC}
\maketitle

%The abstract.
\begin{abstract}
The MAGIC-I telescope is the largest single-dish Imaging Atmospheric 
Cherenkov telescope in the world. A second telescope, MAGIC-II, will 
operate in coincidence with MAGIC-I in stereoscopic mode. MAGIC-II is a 
clone of MAGIC-I, but with a number of significant improvements, namely a 
highly pixelized camera with a wider trigger area, improved optical analog 
signal transmission and a 2-4 GSps fast readout. All the technical elements 
of MAGIC-II were installed by the end of 2008. The telescope is currently 
undergoing commissioning and integration with MAGIC-I. An update of the 
technical performance of MAGIC-I, a description of all the hardware 
elements of MAGIC-II and first results of the combined technical performance 
of the two telescopes will be reported.
\end{abstract}


\section{Introduction}

The 17m diameter MAGIC~\cite{MAGIC} telescope is as of today the largest
single dish Imaging Atmospheric Cherenkov telescope (IACT) for very high
energy gamma ray astronomy with the lowest energy threshold among
existing IACTs. It is installed at the Roque de los Muchachos on the
Canary Island La Palma at 2200 m altitude and has been in scientific
operation since summer 2004. In the past years MAGIC has been
upgraded by the construction of a twin telescope with advanced photon
detectors and readout electronics. The two telescope system,
is designed to achieve an improved sensitivity in
stereoscopic/coincidence operation mode and simultaneously lower the
energy threshold. 

\begin{figure}
\begin{center}
\includegraphics [width=0.5\textwidth]{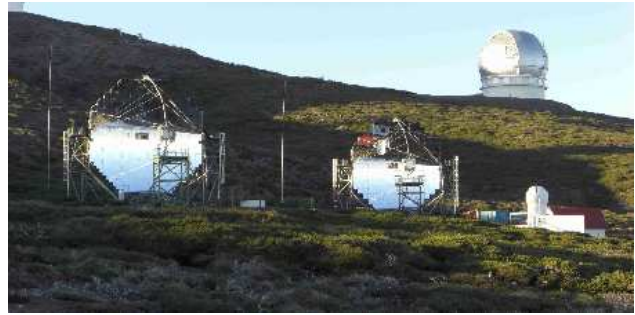}
\end{center}
\caption{The two MAGIC telescopes in April 2009. The first telescope, on the left, 
operates regularly since 2004. The second telescope can be seen on the right. 
The frame, mirrors, active mirror control and drive hardware were  
installed in Summer 2008.}
\label{fig:magic2}
\end{figure}
All aspects of the wide physics program addressed by the MAGIC
collaboration ranging from astrophysics to fundamental physics will
benefit from an increased sensitivity of the instrument. The
expected lower energy threshold of the MAGIC two telescopes will have an impact 
on pulsar studies and extend the accessible redshift range, which is limited by the
absorption of high energy $\gamma$-rays by the extragalactic
background light. Simultaneous observations with the FERMI satellite
will allow detailed studies of the high energy phenomena in the Universe 
in the wide energy range between 100~MeV and 10~TeV.

Detailed Monte Carlo studies have been performed to study the expected
performance of the telescope system~\cite{MAGICII_MC}. In stereo observation mode,
i.e. simultaneously observing air showers with both telescopes, the
shower reconstruction and background rejection power are significantly
improved. This results in an better angular and energy resolution
and a reduced analysis energy threshold. The overall sensitivity is
expected to increase by a factor of 2 over the whole energy range
and foreseably larger below 100~GeV.
Following the results of a dedicated MC study showing moderate
dependence of the sensitivity on the distance of the two telescopes the
second MAGIC telescope has been installed at a distance of 85~m from
the first telescope.

In order to minimize the time and the resources required for design
and production the second MAGIC telescope is in most fundamental
parameters a clone of the first telescope. The lightweight
carbon fiber reinforced plastic telescope frame, the drive 
system~\cite{drive} and the active
mirror control (AMC) are only marginally improved copies of the first
telescope. Both telescopes will be able to reposition within
30-60 seconds to any sky position for fast reaction to GRB alerts.

\begin{figure}
\begin{center}
\includegraphics [width=0.9\columnwidth,angle=270]{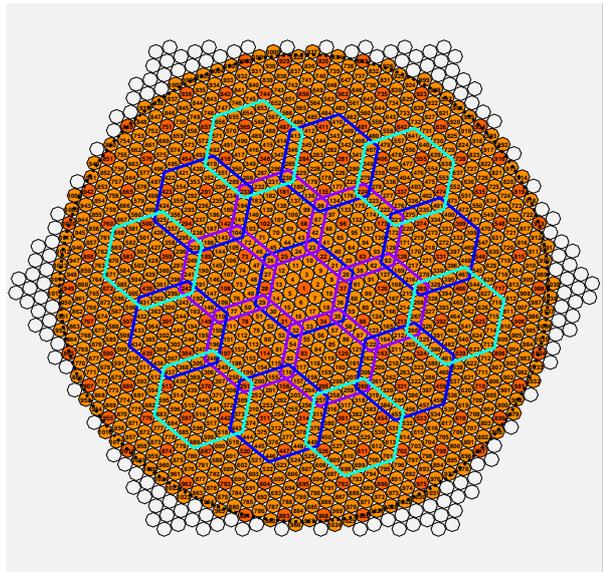}
\caption{A scheme of the MAGIC-II camera. Only colored pixels in a
  round configuration will be equipped. The hexagonal shapes indicate
  the trigger region, which is almost twice as large as the trigger region 
of the first telescope.}
\end{center}
\label{fig:camera}
\end{figure}

Newly developed components are employed whenever they allow cost
reduction, improved reliability or most importantly increased physics
potential of the new telescope with reasonable efforts. Larger 1~m$^2$
mirror elements have been developed for MAGIC-II reducing cost and
installation efforts. The newly developed MAGIC-II readout system
features ultra fast sampling rates and low power consumption. 
In the first phase the camera has been equipped with increased quantum
efficiency (QE) photomultiplier tubes (PMTs), while a modular camera
design allows upgrades with high QE hybrid photo detectors (HPDs). A
uniform camera with 1039 identical 0.1$^o$ field of view (FoV) pixels
(see figure \ref{fig:camera}) allows an increased trigger area compared
to MAGIC-I.

The entire signal chain from the PMTs to the FADCs is designed to have
a total bandwidth as high as 500 MHz. The Cherenkov pulses from
$\gamma$-ray showers are very short (1-3 ns). The parabolic shape of
the reflector of the MAGIC telescope preserves the time structure of
the light pulses. A fast signal chain therefore allows one to minimize
the integration time and thus to reduce the influence of the
background from the light of the night sky (LONS). In addition a
precise measurement of the time structure of the $\gamma$-ray signal
can help to reduce the background due to hadronic background
events~\cite{timing}.

Both telescopes can be seen in figure~\ref{fig:magic2}. The frame of MAGIC-II
and a fraction of the mirrors were installed back in 2007. The remaining
hardware was installed in the Summer 2008. The telescope is currently
undergoing extensive tests and integration with MAGIC-I. The system of two 
telescopes will end its commissioning phase in Fall 2009.

The telescopes have been recently renamed ``MAGIC Florian Goebel Telescopes'' in
memory of the project manager of MAGIC-II, who died shortly before 
completing the telescope in 2008.

In the following the main technical features of the second telescope 
and several upgrades to the first telescope are discussed. References to 
more detailed contributions to the same conference are provided. A 
description of the performance of the two telescope system will be 
presented at the conference.

\section{Mirrors}

Like in MAGIC-I the parabolic tessellated reflector consists of about 250
individually movable 1~m$^2$ mirror units, which are adjusted by the
AMC depending on the orientation of the telescope. While in MAGIC-I
each mirror unit consists of 4 individual spherical mirror tiles mounted on
a panel, MAGIC-II is equipped with 1~m$^2$ spherical mirrors
consisting of one piece. This reduces cost and manpower because 
it is no longer necessary to align all four mirrors individual 
tiles inside one panel before installing the panels at the telescope.

Two different technologies have been used for the production of the
1~m$^2$ mirrors. Out of the 247 mirror tiles, 143 are
all-aluminum mirrors consisting of a sandwich of two 3~mm thick Al
plates and a 65~mm thick Al honeycomb layer in the center. During
production the sandwich is already bent into a spherical shape,
roughly with the final radius of curvature. The polishing of the
mirror surface by diamond milling is done by the LT Ultra
company. Finally, a protecting quartz coating is applied. The
reflectivity $refl$ and the radius $R_{90}$ of the circle containing
90\% of the spot light have been measured to be around $refl$ = 87\%
and $R_{90}$ = 3~mm.

The remaining 104 mirror tiles are produced as a 26~mm tick
sandwich of 2~mm glass plates around a Al honeycomb layer using a cold
slumping technique. The frontal glass surface is coated with a
reflecting Al layer and a protecting quartz coating. The glass-Al
mirrors show a PSF which almost doubles ($\sim$6~mm) that of the all-Al mirrors
but the light spot is still well inside the size of a camera pixel.

\begin{figure}
\begin{center}
\includegraphics [width=0.9\columnwidth]{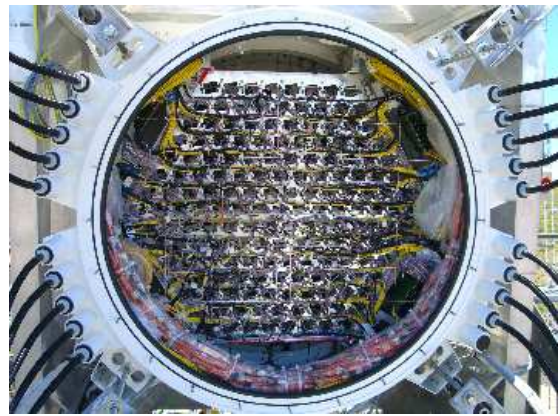}
\caption{A view of the MAGIC-II camera from the back shortly after its physical 
installation at the site and before plugging in the 7-pixel clusters. The 169 
cluster modules are inserted from the camera front into the square holes which 
can be seen on the support metal plate. The pixel signals are transfered to the 
control house through optical fibers bundled inside the outer black cable ducts.}
\end{center}
\label{fig:camera_electronics}
\end{figure}

\section{Camera}

A modular design has been chosen for the camera of the MAGIC-II
telescope~\cite{MAGICII_camera}. Seven pixels in a hexagonal
configuration are grouped to form one cluster, which can easily be
removed and replaced. This allows easy exchange of faulty
clusters. More importantly, it allows full or partial upgrade with
improved photo detectors. The 3.5$^o$ diameter FoV is similar to
that of the MAGIC-I camera. The MAGIC-II camera is uniformly
equipped with 1039 identical 0.1$^o$ FoV pixels in a round
configuration (see figures \ref{fig:camera} and \ref{fig:camera_electronics}).

In the first phase increased QE PMTs have been installed. The Hamamatsu
R10408 6 stage PMTs with hemispherical photocathode typically reach a
peak QE of 34\%. The PMTs have been tested for low
afterpulsing rates, fast signal response ($\sim$1 ns
FWHM) and acceptable aging properties.

Hamamatsu delivers PMT modules which include a socket with a
Cockcroft-Walton type HV generator. The PMT socket and all the
front-end analog electronics is assembled to form a compact pixel
module. The broadband opto-electronic front-end electronics amplifies
the PMT signal and converts it into an optical pulse, which is
transmitted over optical fibers to the counting house. 

A cluster consists of 7 pixel modules and a cluster body which
includes common control electronics, power distribution and a
test-pulse generator.% (see figure~~\ref{fig:cluster}). 
On the front
side the PMTs are equipped with Winston cone type light guides to
minimize the dead area between the PMTs. The slow control electronics
sets the pixel HV and reads the anode currents, the HV values and the
temperature of each pixels. It is in turn controlled by a PC in the
counting house over a custom made RS485 and VME optical link. The camera
control software\cite{MAGICII_slow} is programmed in Labview and can
be remotely steered by a central computer\cite{MAGICII_arehucas}.

The calibration system of MAGIC-II\cite{MAGICII_calibration} 
is based on a frequency tripled passively Q-Switched Nd-YAG laser, 
operating at the third harmonic at 355 nm, which has been installed in the
center of the mirror dish. The pulse width at 355nm is 700ps. For providing 
a large dynamic range we are using two rotating filter wheels under computer 
control that allow one to illuminate the camera with intensities within 
100 steps from single to 1000 photoelectrons. MAGIC-I will be equipped 
with a similar system in the next months.

The flexible cluster design allows field tests of this new technology 
within the MAGIC-II camera without major interference with the rest of the camera.
The first test will in fact take place in the next months:
it is planned to equip six 7-pixel modules in the
outermost ring of MAGIC-II with HPDs~\cite{MAGICII_HPDs} (at the corners of 
the hexagon in figure~\ref{fig:camera} which are not instrumented with PMTs). 
These HPDs feature peak QE values of 50\%. 

The smaller trigger area and somewhat lower light conversion efficiency 
of the first MAGIC telescope will limit the performance of the telescope system.
This justifies a recent decision to upgrade the camera of MAGIC-I. 
The new MAGIC-I camera will be a clone of the camera of MAGIC-II, i.e., 
will have an increased trigger area and will be fully equipped with 
0.1$^o$ FoV pixels. However its inner section (about 400 pixels) 
may be readily equipped with HPDs, i.e. the sensitivity of the new camera would
significantly increase for low energy showers. The camera frame and electronics are 
already under construction. The readout will also be upgraded to a digitizing
system similar to that of MAGIC-II (see below).

\section{Readout}

The optical signals from the camera are converted back to electrical
signals inside the counting house\cite{MAGICII_readout}. 
The electrical signals are split in
two branches. One branch is further amplified and transmitted to the
digitizers while the other branch goes to a discriminator with a
software adjustable threshold. The generated digital signal has a
software controllable width and is sent to the trigger system of the
second telescope with a software adjustable time delay. 
Scalers measure the trigger rates of the individual pixels. 
 
\begin{figure}
\begin{center}
\includegraphics [width=\columnwidth,angle=0]{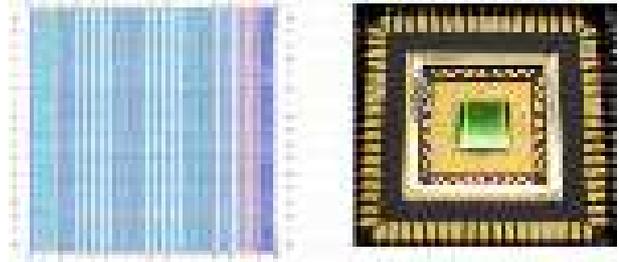}
\caption{Microphotograph of the Domino chip (left) and inside the package (right).}
\label{fig:domino}
\end{center}
\end{figure}

The new 2~GSamples/s digitization and acquisition system is based upon
a low power analog sampler called Domino Ring Sampler (see
figure:~\ref{fig:domino}). The analog signals are stored in a multi
capacitor bank (1024 cell in DRS version 2) that is organized as a ring buffer,
in which the single capacitors are sequentially enabled by a shift
register driven by an internally generated 2 GHz clock locked by a PLL
to a common synchronization signal. Once an external trigger has been
received, the sampled signals in the ring buffer are read out at a
lower frequency of 40~MHz and digitized with a 12 bits resolution
ADC. 

Data management is performed by 9U VME digital boards which handle the
data compression and reformatting as well. Every board hosts 80 analog
channels plus auxiliary digital signals for trigger and monitor
purposes.  For a 1~kHz trigger rate and a 2 GHz frequency sampling, the
data throughput can be as high as 100 MBytes/s thus being a challenge for
modern data transmission and storage solutions. The data are
transferred to PCI memory via Gbit optical links using the CERN S-link
protocol and to the mass storage system\cite{MAGICII_storage}. 

The MAGIC I telescope produces currently per year ~100TByte of raw data 
that is calibrated and reduced on-site. Since 2007 most of the the data 
has been stored and further processed at the official MAGIC-II datacenter at 
PIC, Barcelona\cite{MAGICII_datacenter}. This datacenter is currently 
undergoing an upgrade to accomodate the even larger storage demands of MAGIC-II.

The newest version of the DRS chip (DRS version 4) features a number of
advantages over DRS-2, so work is underway to upgrade the system in the
next year to DRS-4.

\section{Trigger}
 
The trigger system of the second telescope like the trigger of the
first telescope\cite{trigger} is based on a compact next neighbor logic. 
However, the uniform camera design allows an increased trigger
area of 2.5$^o$ diameter FoV. This increases the potential to study
extended sources and to perform sky scans. 

When the two telescopes are operated in stereo mode a coincidence
trigger (so-called ``level 3'' trigger) between the two telescopes 
rejects events which only triggered one telescope. In order to 
minimize the coincidence gate in the level 3 trigger, the triggers
produced by the individual telescopes will be delayed in a time 
which depends of the geometry of the telescopes. This will reduce 
the overall trigger rate to a rate which is manageable by the data 
acquisition system. 

Starting in 2007, an additional trigger runs in parallel with the standard
next neighbour trigger in first telescope. This so-called ``sumtrigger'' 
\cite{sumtrigger} operates on the analog sum of groups of 18 pixels and 
has allowed to lower the trigger threshold of the MAGIC telescope 
by a factor of two to 25 GeV.

\section{Acknowledgments}

We would like to thank the IAC for excellent working conditions. The
support of the German BMBF and MPG, the Italian INFN and the Spanish
MICINN, the Swiss ETH and the Polish MNiI is gratefully acknowledged.

\end{document}